\documentclass[twocolumn,prb,showpacs,aps]{revtex4}

\usepackage{graphicx}
\usepackage{dcolumn}
\usepackage{bm}
\usepackage{indentfirst}

\usepackage{epstopdf}

\begin{document}

\title{Superconductivity-Enhanced Bias Spectroscopy in Carbon Nanotube Quantum Dots}

\author{K. Grove-Rasmussen}
\altaffiliation{Present address: NTT Corporation, NTT Basic Research Laboratories, 3-1 Morinosato Wakamiya, Atsugi-shi, 243-0198 Kanagawa, Japan}
\email{grove@will.brl.ntt.co.jp}
\author{H. I. J\o rgensen}
\author{B. M. Andersen}
\author{J. Paaske}
\author{T. S. Jespersen}
\author{J. Nyg\aa rd}
\author{K. Flensberg}
\author{P. E. Lindelof}

\affiliation{Nano-Science Center, Niels Bohr Institute,
University of Copenhagen, Universitetsparken 5, 2100~Copenhagen \O ,
Denmark}
\date{\today}
\addtolength{\textheight}{+5mm}
\begin{abstract}
We study low-temperature transport through carbon nanotube quantum
dots in the Coulomb blockade regime coupled to niobium-based superconducting leads. We observe pronounced conductance peaks at finite source-drain bias, which we ascribe to elastic and
inelastic cotunneling processes enhanced by the coherence peaks in
the density of states (DOS) of the superconducting leads. The inelastic
cotunneling thresholds display a marked dependence on
gate voltage caused by different tunneling-renormalizations
of the two subbands in the nanotube. Finally, we discuss the gate-dependent sub-gap structure observed in a strongly coupled device with odd electron occupation.
\end{abstract}


\maketitle

Superconducting electrodes provide a useful means of sharpening the
spectroscopic features observed in tunneling experiments. In the
superconducting phase, an otherwise nearly constant DOS acquires a gap of width $2\Delta$ centered at the
Fermi level and characteristic sharp coherence peaks at the
gap-edges $\pm \Delta$. These peaks
transform a featureless metallic electrode into a
high-resolution tunneling probe. This widely used investigative tool~\cite{Wolf1985} was demonstrated already by
Giaever's seminal work~\cite{GiaeverSS1960} from 1960 and more recently
used to obtain a high resolution bias-spectrum of
the levels in a metallic Al-nanoparticle \cite{Ralph}.
\linebreak
\indent
Here we report low-temperature transport measurements in which this
type of BCS-focusing promotes an otherwise featuresless elastic
cotunneling conductance to sharp peaks at bias voltages $V_{sd}=\pm
2\Delta/e$, corresponding to the onset of quasiparticle cotunneling.
In the same way, inelastic cotunneling processes involving
transitions between two subbands in the nanotube are revealed as
sharp peaks rather than steps or cusps in the nonlinear conductance.
This sharpening of cotunneling lines inside the Coulomb diamonds
allows us to investigate more closely the tunneling-induced gate
voltage dependence of the orbital splitting \cite{Holm}. Finally, we
discuss an unusual sub-gap structure observed in a particularly well
coupled device, signalling the importance of both multiple Andreev
reflections (MAR) and dynamically generated bound states in spinful
dots.
\linebreak
\indent
A number of experiments have already investigated interesting
aspects of quantum dots with superconducting electrodes, such as
supercurrent \cite{Kasumov,Hakonen2007PhRvL,strunk}, MAR
\cite{MorpurgoScience}, and effects of size and charge quantization
in the Fabry-Perot
\cite{Jarillosupercurrent,hij,wernsdorfer,LiuRiverside}, Kondo
\cite{BuitelaarKondo,Cleuziou,kgr,Tarucha2007PhRvL,Eichler,tsj}, and
Coulomb blockade regimes
\cite{Buitelaar,Cleuziou,hijpi,vanJordan,doh}. We present
measurements performed on two different single walled carbon
nanotube (SWCNT) quantum dots coupled to niobium-based
superconducting leads \cite{MorpurgoScience,strunk}. Our first
device (device A) is poorly coupled to the leads and shows regular
Coulomb blockade diamonds with clear onset of quasiparticle (elastic
cotunneling) current. The second device (device B) is more strongly
coupled to the leads and displays a four-fold degenerate shell
structure with both elastic and inelastic cotunneling lines which
are sharpened by the superconducting leads.
\linebreak
\indent
\begin{figure}[t!]
\begin{center}
\includegraphics[width=0.46\textwidth]{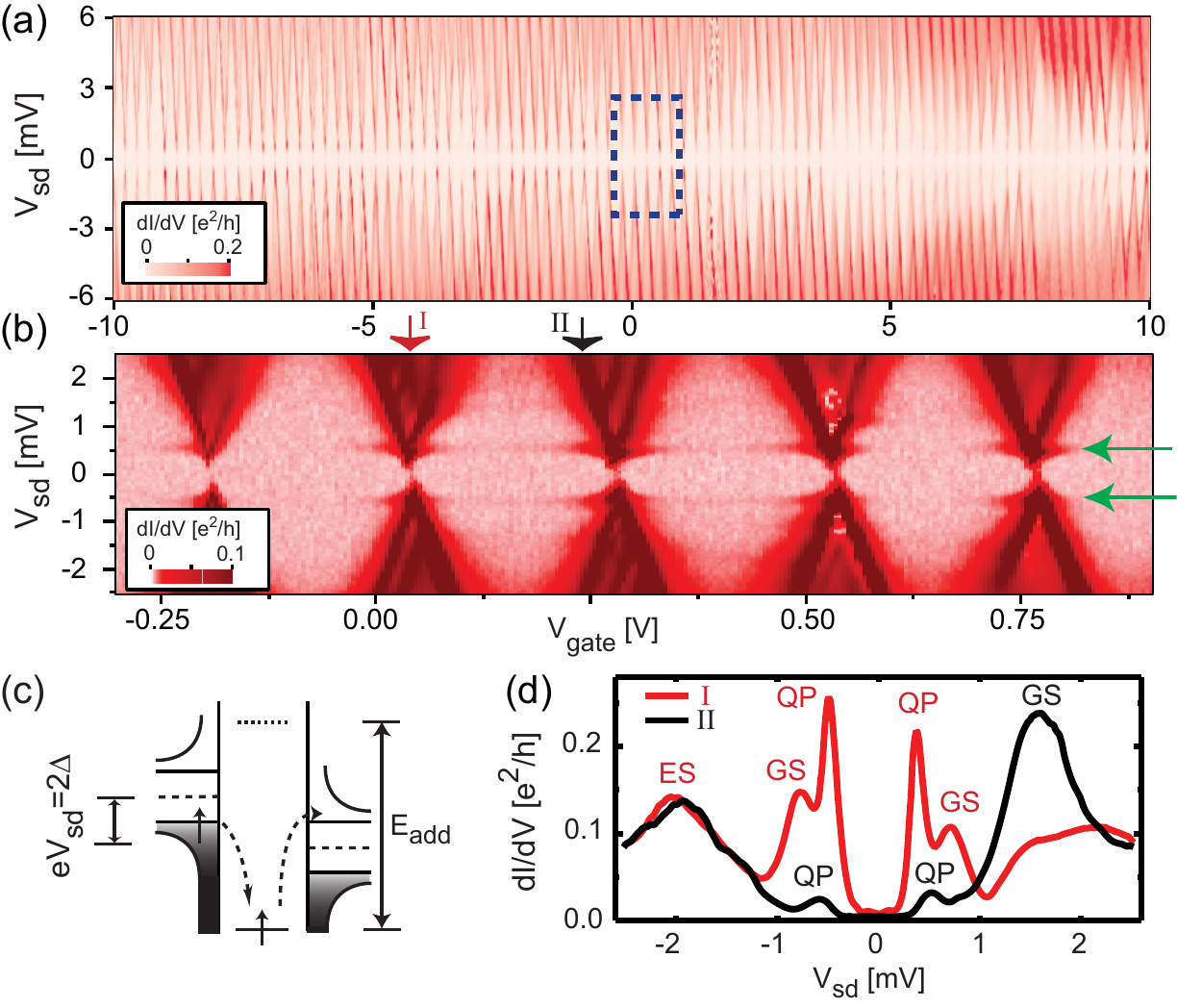}
\end{center}
\caption{(Color online) Device A. (a) Bias spectroscopy at
$T=0.3$\,K of a SWCNT coupled to Ti/Nb/Ti leads. (b)
Conductance within the dashed rectangle in (a). The
onset of quasiparticle cotunneling is clearly seen at $eV_{sd}
=\pm 2\Delta$ (green arrows). (c) Schematic energy diagram for a
viable cotunneling process in the center of a diamond. (d) Bias cuts
at positions (I,II) in (b) showing conductance peaks due to
quasiparticle (QP) cotunneling, and sequential tunneling into ground
(GS) and excited states (ES), respectively.\label{figure1}}
\end{figure}
The devices are made by growing high quality SWCNT
by chemical vapor deposition from predefined catalyst
islands \cite{hij,kgr}. Contacts are made of Nb-based trilayers, $x$/Nb/$x$ (about 5/60/10\,nm), with $x$=Pd,Ti where $x$ is thermally evaporated and the Nb is deposited by sputtering technique. The superconducting trilayers are tested
via four terminal devices on the same chip and show
transition temperatures close to $T=9$\,K. Nevertheless,
the actual gap at the nanotube indicates a critical temperature of
$T_c\approx 1.7$\,K, which might be related to the formation of
NbO ($T_c \sim 1.4$\,K) \cite{NbO} or contamination of the lower Nb/x interface. Samples from three different processing rounds revealed similarly reduced $T_c$ in transport measurements. However, the high quality of the presented measurements is a promising first step towards Nb based SWCNT Josephson junctions.
\linebreak
\indent
Figure \ref{figure1}(a) shows the conductance versus gate, and bias voltage (bias spectroscopy plot) at $T=0.3$\,K
for device A consisting of a carbon nanotube quantum dot
coupled weakly to Ti/Nb/Ti leads.  It reveals more than 80 regular Coulomb
diamonds, illustrating that only one quantum dot is defined in this high quality carbon nanotube. The
charging energy, $U \approx 5-6$\,meV is estimated from the height
of the diamonds and no clear shell structure is observed\footnote{A variation of the addition energies is seen, but no clear four-electron shell structure is observed as is the case of device B. The origin of this difference is not understood, except that the energy scales related to device B are much larger than in device A, making such effects more visible.}. A region of highly
suppressed conductance around zero bias is clearly observed
for all gate voltages reflecting the superconducting energy gap of
the leads. Figure \ref{figure1}(b) shows the Coulomb diamonds in the
dashed rectangle of Fig.\ \ref{figure1}(a). In Coulomb blockade (inside the diamonds)
the onset of quasiparticle tunneling is seen as horizontal lines (conductance ridges) at $eV_{sd} =\pm
2\Delta \approx \pm 0.55$\,meV (green arrows). Higher order Andreev
reflections, which would lead to current below the gap, are strongly suppressed due to the poor coupling to the
leads. Inside the diamonds, the onset of quasiparticle
cotunneling corresponds to an alignment of the superconducting
DOS peaks and involves elastic cotunneling processes as depicted in Fig.\ \ref{figure1}(c). At the charge degeneracy
points, conductance inside the gap is due to Andreev
reflections \cite{Buitelaar}. Figure \ref{figure1}(d) shows a bias
cut slightly off resonance (I) and further off resonance (II). The onset of
quasiparticle tunneling at (I) involves a sequential tunneling
process and is therefore much stronger than in (II). Peaks at higher
bias are due to sequential tunneling to ground and
excited states, respectively.
\begin{figure}[t!]
\begin{center}
\includegraphics[width=0.46\textwidth]{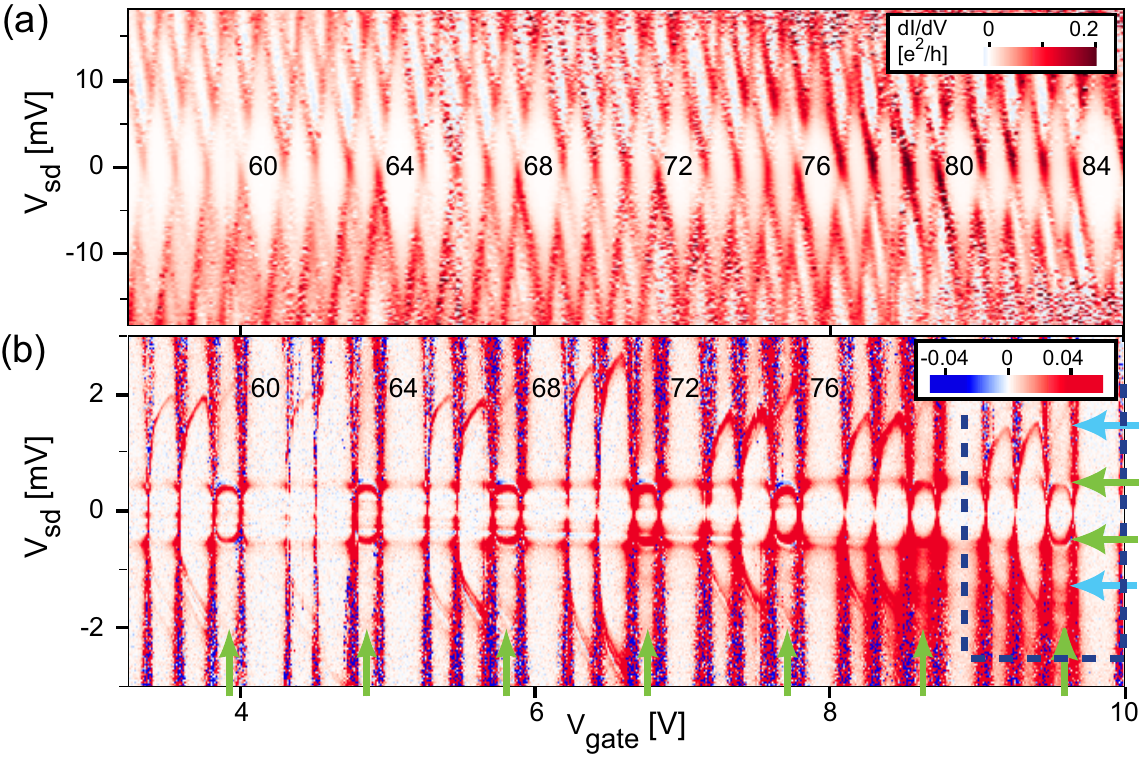}
\end{center}
\caption{(Color online) Device B. (a) Bias spectroscopy at
$T=6.5$\,K, i.e. above $T_c\approx 1.7$\,K. Four-fold shell structure
is observed with occupation numbers written in the large Coulomb
diamonds corresponding to a full shell (zero occupation does not
correspond to half-filling). (b) Same gate region restricted to
$|V_{sd}|\leq 3$\,mV at $T=0.3$\,K showing inelastic cotunneling
features (blue arrows) as well as the onset of elastic quasiparticle
cotunneling (horizontal green arrows). Every fourth diamond shows pronounced gate-dependent
conductance peaks near $|V_{sd}|=2\Delta/e$ (vertical green arrow).
\label{figure2}}
\end{figure}

\begin{figure}[t!]
\begin{center}
\includegraphics[width=0.46\textwidth]{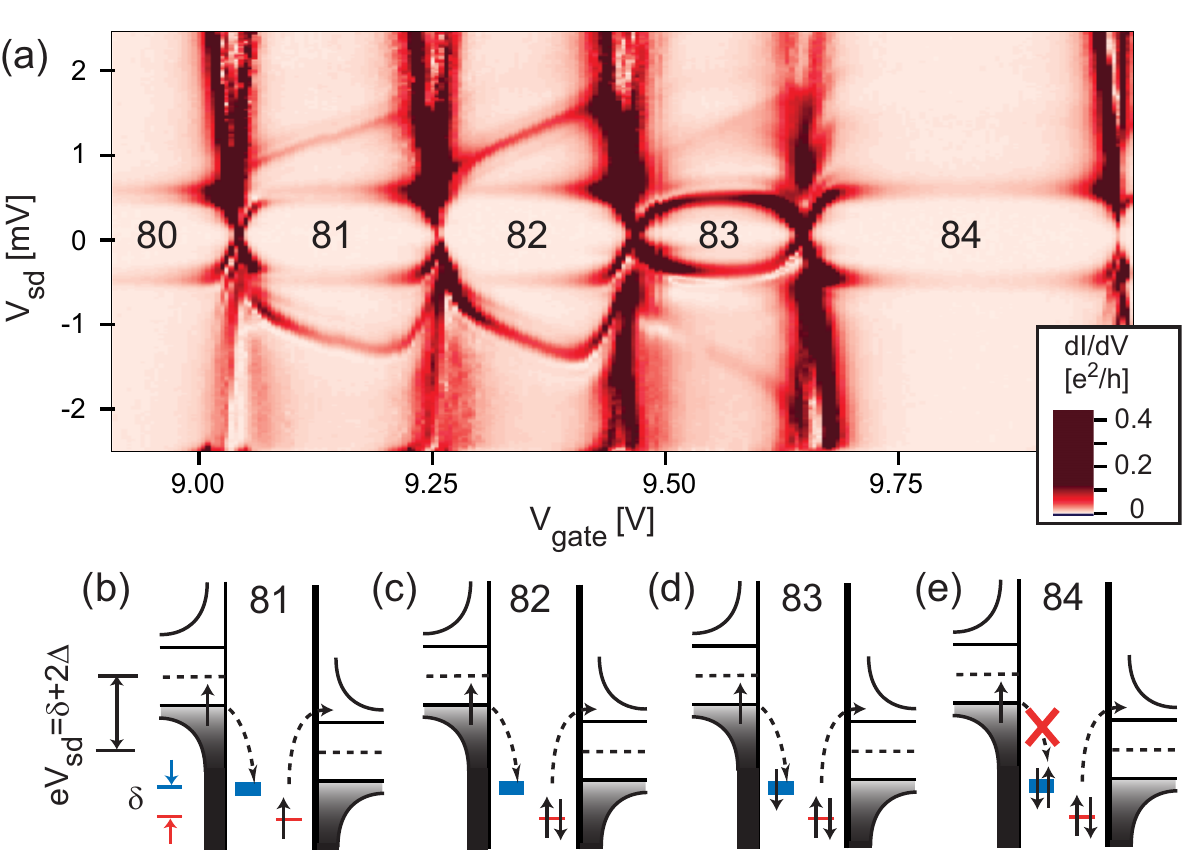}
\end{center}
\caption{(Color online) Device B. (a) Detailed bias spectroscopy at
$T=0.3$\,K of one four-electron
shell [dashed blue
rectangle in Fig.\ \ref{figure2}(b)]. (b-e) Schematic energy
diagrams illustrating inelastic cotunneling processes relevant for
fillings of 1-4 electrons in a shell, e.g., corresponding to charge states 81-84. The thick blue (thin red) level represents the strongly (weakly) coupled
orbital, which both are coupled more weakly to the right electrode (drain) shown by a thicker barrier.\label{figure3}}
\end{figure}

Figure \ref{figure2}(a) shows a bias spectroscopy plot for the
second, better coupled, device (device B) at $T=6.5$\,K well
above the transition temperature ($T_c\approx 1.7$\,K) of the superconducting Pd/Nb/Pd
layer. A regular pattern is seen with three consecutive
small Coulomb diamonds followed by a larger diamond,
reflecting the filling of shells consisting of two nearly degenerate
orbitals. Numbers in Fig.\ \ref{figure2} denote the additional electron number on the
SWCNT for filled shells counted from $V_{gate} \approx -10$\,V.
A charging energy of $U \approx 12$\,meV and a level spacing of $\Delta E \approx 6$\,meV
are found from the plot. Figure \ref{figure2}(b) shows the conductance at
$T=0.3$\,K, i.e.\ below $T_c$. The lines
at $V_{sd} \approx \pm 0.55$\,mV (green horizontal arrows) are caused by elastic quasiparticle
cotunneling as illustrated by Fig.\ \ref{figure1}(c).
\linebreak
\indent
Inelastic cotunneling lines [Fig.\ \ref{figure2}(b), blue arrows]
are observed at higher $V_{sd}$ for electron numbers $N+1$, $N+2$,
$N+3$, but not in the full shell ($N+4$ electrons). These lines have
a marked gate voltage dependence which resembles the 'double-headed
arrow' structure pointed out in Ref.\ \onlinecite{Holm}. With the
enhanced spectroscopy offered by the superconducting leads we also
observe an additional third 'arrowhead' outside the strong elastic
cotunneling lines appearing in every fourth (N+3) diamond, i.e. a
possible Kondo ridge at this temperature in the normal state. The
high $T_c$ and critical field of the Nb-films prevented us from
confirming the presence of a normal-state Kondo resonance, insofar
as this resonance would already be suppressed by the magnetic field.
The inelastic cotunneling lines are seen more clearly in Fig.\
\ref{figure3}(a) which shows detailed measurements from the dashed
rectangle in Fig.\ 2(b).  The cotunneling processes are depicted in
Fig.\ \ref{figure3}(b-e), involving a weaker coupled orbital
(orbital 1) [thin red line in Fig.\ \ref{figure3}(b)] and a stronger
coupled orbital (orbital 2) [thick blue line in Fig.\
\ref{figure3}(b)] split by $\delta$. Such processes are allowed for
all but the charge state corresponding to a filled shell [Fig.\
\ref{figure3}(e)], consistent with Fig.\ \ref{figure3}(a). As
demonstrated by Holm {\it et al.} \cite{Holm}, a difference in
tunnel-couplings to the two orbitals in the quantum dot, gives rise
to a gate-dependence of the threshold for inelastic cotunneling.
With superconducting leads, such tunneling renormalization produces
a gate-dependent shift of the unrenormalized threshold at $eV_{sd}
\geq \delta+2\Delta$, as observed.

\begin{figure}[t!]
\begin{center}
\includegraphics[width=0.44\textwidth]{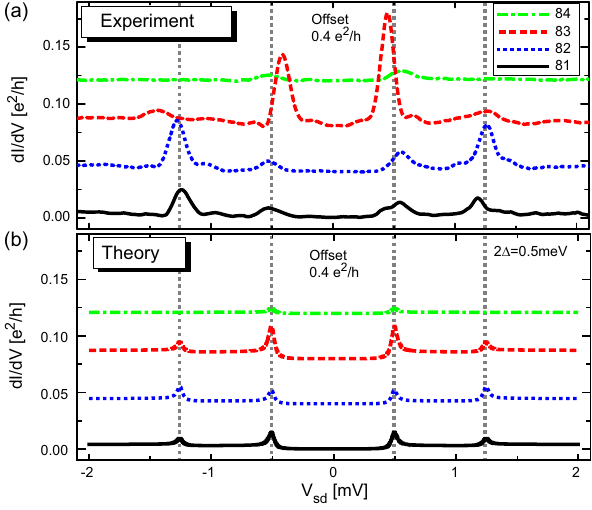}
\end{center}
\caption{(Color online) Comparison of measured (a) and calculated (b) conductance versus $V_{sd}$ in the center of diamonds 81-84 in Fig.\ \ref{figure3}(a). In the calculation $\delta=3\Delta$, $k_B T=0.03$\,meV and addition energies [12.6, 12.6, 10.8 \& 20.7]\,meV extracted from the width of the diamonds 81-84 have been used. \label{theoplot}}
\end{figure}

In Fig. \ref{theoplot}(a) we show conductance line-cuts through the center of each diamond 81-84 in Fig.  \ref{figure3}(a).
The variation of the peak heights can be understood from the number
of elastic (EL) and inelastic (INEL) cotunneling channels in
each diamond, as summarized in Table \ref{tab1}.
\begin{table}[b!]
\caption{Number of cotunneling channels.\label{tab1}}
\begin{ruledtabular}
\begin{tabular}{ccccc}
Shell filling & 1 & 2 & 3 & 4 \\ \hline
  \mbox{EL} & $\left( \begin{array}{c} 2 \\ 4 \end{array} \right)$ & $\left( \begin{array}{c} 2 \\ 2 \end{array} \right)$ & $\left( \begin{array}{c} 4 \\ 2 \end{array} \right)$ & $\left( \begin{array}{c} 2 \\ 2 \end{array} \right)$ \\
 \mbox{INEL} & 2 & 4 & 2 & 0\nonumber
\end{tabular}
\end{ruledtabular}
\end{table}
There, the EL notation indicates whether the tunneling takes place through
the higher- or lower-energy orbital. For
example, in case of filling 1 [see Fig.\ \ref{figure3}(b)], a total
of 6 EL channels contribute: 2 from the upper and 4 from the lower orbital.

In Fig.\ \ref{theoplot}(b) we show the results of a calculation of
the lowest order nonlinear cotunneling conductance for the four
different charge states in a single shell. The calculation involves
the quasiparticle tunneling rates $W^{\alpha\beta}_{ij}$ between
leads $\alpha,\beta=L,R$ and orbitals $i,j=1,2$. For example, the
rate for the process shown in Fig.\ \ref{figure3}(b) is
\begin{equation}
W^{LR}_{12}\!=\!\frac{e}{h}\frac{4}{U^2}\!\!\int_{-\infty}^\infty\!\!\!\!\!dE\,\Gamma^{L}_{2}(E)\Gamma^{R}_{1}(E+\delta)
f_L(E) [1-f_R(E+\delta)]
\end{equation}
with $f_\alpha(E)=f(E-\mu_\alpha)$ the Fermi function, $\Gamma^{\alpha}_{i}(E)=\Gamma^{\alpha}_{i}
|E-\mu_\alpha|/\sqrt{(E-\mu_\alpha)^2-\Delta^2}$, and $\Gamma^{\alpha}_{i}=\pi\nu_{F}|t_{\alpha,i}|^{2}$ in terms of the tunneling amplitudes $t_{\alpha,i}$. Solving the steady-state rate equations we obtain the orbital occupation numbers as a function of $V_{sd}$ and the current is readily determined. In agreement with experiment one sees from Fig.\ \ref{theoplot} that the inelastic cotunneling peak is largest for diamond 82 due to the larger number of tunneling possibilities. Likewise the elastic $2\Delta$-peak is largest for diamond 83 due to the larger coupling of orbital 2 to the leads. The calculation does not reproduce the small shifts of some of the peaks seen in the experimental data, as well as the ratio between the amplitude of the elastic and inelastic peaks in diamond 81. We speculate that the latter discrepancy will be  removed by including higher order processes involving the strongly coupled orbital 2.  The width of the peaks arise from a small smearing factor ($\eta=0.01\mbox{meV}$) used in the DOS of the superconducting leads mimicking a constant inelastic scattering rate in the contacts. We ascribe the larger width of the measured peaks to tunnel-broadening of the excited states which is not included in the calculation.

For the results shown in Fig.\ \ref{theoplot} we have used
$\Gamma_1^R=0.07$\,meV, $\Gamma_2^R=0.1$\,meV, $\Gamma_1^L =
1.0$\,meV, and $\Gamma_2^L = 1.8$\,meV, yielding an asymmetry factor
in the order of $\sum_i \Gamma^L_i/\sum_i\Gamma^R_i \approx 15$
consistent with an upper bound\footnote{The Coulomb peaks measured at $T=6.5$K are already slightly suppressed by temperature.} of approximately 40 extracted from
the Coulomb peak heights in the shell at
6.5K.
The couplings are estimated from the sequential current at large
positive ($I^+$) and negative bias ($I^-$) at the charge degeneracy
point for adding the first electron in a four-fold degenerate shell.
These currents are given by
$I^{+/-}=\frac{e}{h}\frac{2\Gamma_1^{L/R} \Gamma_2^{L/R}
(\Gamma_1^{R/L}+\Gamma_2^{R/L})}{2(\Gamma_2^{L/R}
\Gamma_1^{R/L}+\Gamma_1^{L/R} \Gamma_2^{R/L})+\Gamma_1^{L/R}
\Gamma_2^{L/R}} \approx 11$\,nA/$-3.5$\,nA
in reasonable agreement with the experiment at the Coulomb resonance
involving electron charge states 80 and 81, $I^{+/-}_{exp} \approx
12$\,nA/$-3$\,nA. Moreover, the chosen couplings lead to a
gate-voltage slope \cite{Holm} of the inelastic $2\Delta+\delta$
line of $d \delta/d (e \gamma
V_g)=4\sum_\alpha(\Gamma^\alpha_2-\Gamma^\alpha_1)/\pi
U\approx0.084$ (average $U=12.0$\,meV), which agrees with the
experimental result of $0.082$ [average slope of the cotunneling
lines in diamonds 81-83 in Fig.\ \ref{figure3}(a)]. We note that
since $U \gg \Delta$, the expression for the renormalization
obtained in the normal state \cite{Holm} remains valid in the case
of superconducting leads. Based on the couplings
$\Gamma^{\alpha}_{i}$ used above, we estimate the Kondo energies
(temperatures) in diamonds 81, and 83 to be $k_B T_{K,81} \approx
0.2$\,$\mu$eV (2\,mK) and $k_B T_{K,83}\approx 0.03$\,meV (300\,mK),
i.e. much smaller than $\Delta\sim 0.28$\,meV, consistent with our
observation of sub-gap structure rather than an enhanced zero-bias
conductance peak~\cite{BuitelaarKondo}.
\linebreak
\indent
\begin{figure}[t]
\begin{center}
\includegraphics[width=0.46\textwidth]{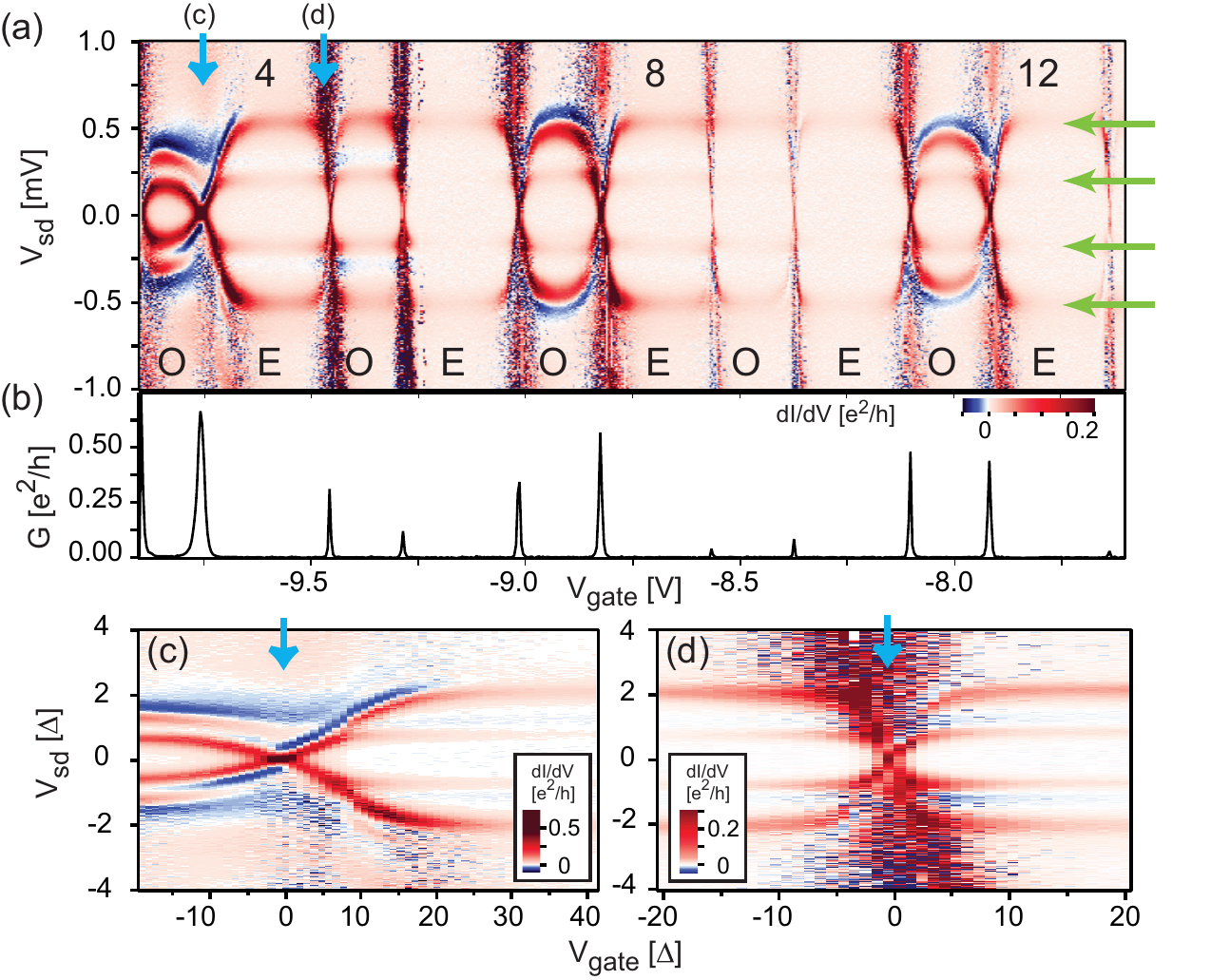}
\end{center}
\caption{(Color online) Device B. (a) Bias spectroscopy at
$T=0.3$\,K for a strongly coupled gate-region with pronounced change
of the sub-gap structure in every fourth diamond. Numbers indicate
the additional electron number in correspondence with Fig.\
\ref{figure2}, while the large/small diamonds are assigned even
(E)/odd (O) electron filling. The horizontal green arrows indicate
the elastic quasiparticle cotunneling lines at $eV_{sd}=\pm
2\Delta\approx \pm 0.5$\,meV together with the lower-lying lines
corresponding to a single Andreev reflection. (b) Linear conductance
(zero bias) versus gate voltage hinting that well and poorly coupled
orbitals are filled consecutively. (c-d) Zoom at the gap structure
around two Coulomb blockade resonances marked by arrows in (a) with
the gate voltage in units of $\Delta$. \label{figure4}}
\end{figure}
\linebreak
\indent
We now discuss measurements from a better coupled gate voltage
region of device B, exhibiting a characteristic rounding of the
elastic quasiparticle cotunneling and an unusual sub-gap conductance
as seen in Fig.~\ref{figure4}(a). For charge states with three
electrons in a shell, the sub-gap structure is especially pronounced
and gate-dependent, indicating that this orbital is particularly
well coupled to the leads. This is supported by the linear
conductance data presented in Fig.~\ref{figure4}(b), showing broad
resonances where the effect is largest (diamonds 3,7 and 11). In
diamond 3, the sub-gap conductance even exceeds the elastic
cotunneling peak in the other diamonds at $eV_{sd}\sim \pm 2\Delta$~\cite{tsj,Eichler}, with
strong peaks at voltages which are different from the expected MAR
positions at $\pm 2\Delta/n$. Figure~\ref{figure4}(a) also displays
marked negative differential conductance, seen as blue lines
reflecting a local minimum in the current at $eV_{sd}\sim \pm 2\Delta$ in
the center of diamonds 3, 7, and 11. A similar negative differential conductance effect has been
reported in a nanotube quantum dot with very different couplings to
source, and drain electrodes~\cite{Eichler}.

In Figs.~\ref{figure4}(c,d) we show a zoom-in of two neighboring
charge degeneracy points ($N=3/4$ and $N=4/5$) marked by vertical
(blue) arrows in Fig.~\ref{figure4}(a). Panel (d) shows a close
resemblance to the behavior expected from MAR in the presence of a
resonant level ($\Gamma\lesssim
\Delta$)~\cite{Yeyati1997PhRvB,Johansson1999PhRvB,Buitelaar} with
the rounding of the sub-gap structure extending some 4$\Delta$ into
the Coulomb blockade diamond. The data in panel (c), on the other
hand, exhibit a qualitatively different sub-gap structure with
pronounced negative differential conductance and a much stronger gate dependence which levels off
at roughly 15-20$\Delta$ away from the charge-degeneracy point.

We speculate that the unusual sub-gap features observed in odd
occupied dots are caused by an interplay between MAR and quantum
Shiba states~\cite{Soda67}, present for spinful dots with
$k_BT_{K} <\Delta$.
For strongly asymmetric couplings, these spin-induced bound
states remain pinned to the stronger coupled lead, at energies
inside the gap given roughly by the exchange coupling with this
electrode. Therefore, new conductance peaks away from the usual $\pm
2\Delta/n$ occur naturally in this scenario, and a bias-scan with
the coherence peaks of the weaker coupled lead gives rise to negative differential conductance at
$\pm 2\Delta$ because of spectral weight transfer from the coherence
peaks of the stronger coupled lead to the bound states.
\linebreak
\indent

In summary, we have demonstrated how superconducting electrodes lead
to dramatic enhancement of cotunneling spectroscopy in carbon
nanotube quantum dots. This revealed pronounced inelastic
cotunneling lines with marked gate-dependence caused by
tunneling-induced level-shifts. Moreover, we discussed the presence
of negative differential conductance and unusual sub-gap conductance in strongly coupled odd
occupied diamonds. Further studies are required to fully uncover the
interesting interplay between MAR and spin correlations in quantum
dots.
\linebreak
\indent
We acknowledge experimental help from J\o rn Bindslev Hansen. This
work was supported by V. K. Rasmussen Foundation, Danish Agency for
Science, Technology and Innovation, Carlsberg Foundation, CARDEQ and
SECOQC projects, and the Danish Research Council.
\bibliography{text}
\end{document}